\documentclass{article}

\usepackage[english]{babel}
\usepackage{verbatim}

\usepackage[letterpaper,top=2cm,bottom=2cm,left=3cm,right=3cm,marginparwidth=1.75cm]{geometry}

\usepackage{amssymb}
\usepackage{graphicx}
\usepackage[colorlinks=true, allcolors=blue]{hyperref}

\title{The System Model and the User Model: \\Exploring AI Dashboard Design}
\author{Fernanda Vi\'egas and Martin Wattenberg\footnote{Both authors are at Harvard University}}

\begin{document}
\maketitle

\begin{abstract}
This is a speculative essay on interface design and artificial intelligence. Recently there has been a surge of attention to chatbots based on large language models, including widely reported unsavory interactions. We contend that part of the problem is that text is not all you need: sophisticated AI systems should have dashboards, just like all other complicated devices. Assuming the hypothesis that AI systems based on neural networks will contain interpretable models of aspects of the world around them, we discuss what data such dashboards might display. We conjecture that, for many systems, the two most important models will be of the user and of the system itself. We call these the \textit{System Model} and \textit{User Model}. We argue that, for usability and safety, interfaces to dialogue-based AI systems should have a parallel display based on the state of the System Model and the User Model. Finding ways to identify, interpret, and display these two models should be a core part of interface research for AI.

\end{abstract}

\section{Introduction}

Cars have gas gauges. Ovens have thermometers. Coffeemakers have blinking lights. Our mechanical devices constantly tell us about their internal state. And for good reason: knowing what's happening under the hood let us use machines safely and reliably.

In this note we contend that AI systems, even ones capable of expressive language, need instrumentation too. Effective human-AI interaction will require more than just conversation, and would benefit from dashboards that report in real-time on the system's internal state. These metaphorical meters and dials will likely be application-dependent, but we suggest certain types of information may be universally important.

\begin{figure}[h]
\centering
\includegraphics[width=1\linewidth]{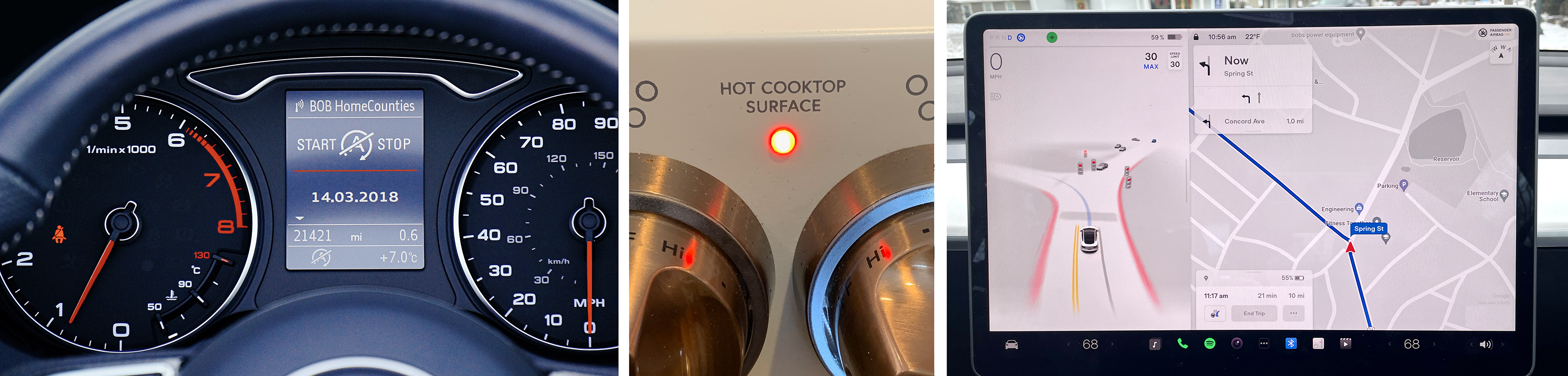}
    \caption{Examples of existing machine instrumentation. Left: tachometer and speedometer in car dashboard. Center: stove top indicator of heat. Right: self-driving display in a Tesla car.}
    \label{fig:dashboards}
 \end{figure}
 
Our argument is based on two conjectures. First, we believe that neural networks contain interpretable models of the world they interact with. Second, we hypothesize that these world models are natural targets for visualizations and other interface elements. In other words, simplified data on the state of these models can be immensely helpful to users, just as data on speed is useful when driving a car, or heat when using an oven.

A real-world example of displaying a world model is the touchscreen of a Tesla car, which shows the inferred state of the road ahead. This view is helpful in letting drivers calibrate trust in the system. In the case of the Tesla, however, the ``world models'' were explicitly designed into the system by its engineers. Our contention is that even when such models are not explicitly built in, neural networks contain them anyway. Once we learn to surface that information and to read it, these models will be just as valuable for user interfaces. Identifying world models is therefore not just an abstract intellectual exercise, but should be seen as a core piece of user interface work.

The Tesla display is obviously tuned for the case of driving a car. For other applications the most helpful type of instrumentation will depend on context. However, we propose that two particular elements of a neural network's world model will play an important role across many different contexts. We call these the \textit{System Model}, which is our term for network's model of itself, and the \textit{User Model}, the network's model of the user interacting with it. Our contention is that many future AI systems should have prominently accessible monitors that show information about the state of both the System Model and the User Model.

To be clear, this argument is tentative and untested. It is meant to spark discussion of a targeted research program in human/AI interaction. We have not run user studies, or even found specific evidence of an System Model or User Model in any large system. Nonetheless, we hope to make the case that this set of ideas warrants further investigation, and may provide a natural framework for designing and investigating user interfaces for AI systems. 

\section{The interpretable world model hypothesis}

Neural networks are often referred to as ``black boxes,'' or opaque systems that defy interpretation. However, there is increasing evidence that systems trained on very basic tasks, such as completing sequences, can develop \textbf{world models}: interpretable representations of aspects of the ``world'' they have been trained on. 

In this speculative note, we want to explore the design implications of what we call the \textbf{world model hypothesis}, by which we mean the idea that the important aspects of a neural network's behavior can be tied to an underlying interpretable model of some element of its world. To be sure, this is a  controversial point. The opposite view--that language models, e.g., are just a ``haphazard'' collection of statistics~\cite{bender2021dangers}--is certainly plausible. Below, we briefly sketch some of the reasons to think the world model hypothesis might hold, in at least enough generality to have implications for interface design. However, a full review of this question is far beyond the scope of this note, and we don't regard the issue as settled ourselves. We think that, even if the concept of ``world model'' doesn't fully explain the internal state of these systems, it is still a useful hypothesis with implications that are worth exploring. We ask the reader to view subsequent sections through that lens.

\subsection{Why the world model hypothesis is plausible}

The main reason to believe the world model hypothesis is that when people look for interpretable representations, they often find them. Sometimes these representations are in plain sight, such as individual neurons that represent salient high-level human concepts. We see such neurons in vision networks, where neuron activations can encode concepts ranging from the presence of curves to lamps to floppy ears~\cite{bau2020understanding, olah2018building}. In natural language processing, high-level concepts such as sentiment have been related directly to individual neurons~\cite{donnelly2019interpretability}.

More generally, one might look for interpretable models that take more complex forms. The technique of ``probing''~\cite{alain2016understanding, belinkov2022probing, hewitt2019structural} can uncover features encoded by arbitrary directions in activation space, or (for nonlinear probes) more exotic geometric forms. For example, such probes reveal that various forms of human-understandable syntactic information seem to be encoded in many NLP-focused neural networks~\cite{hewitt2019structural, chi2020finding, tenney2019bert}. A related technique, TCAV~\cite{kim2018interpretability} has shown success in uncovering concepts defined by sets of examples. Probes have also uncovered models of semantic distinctions as well, including word senses~\cite{reif2019visualizing} and color spaces~\cite{abdou2021can}.

Recently we've also seen evidence that language models can effectively track state and learn in context, and researchers are beginning to identify models and circuits that explain these abilities. At a low level, investigators have uncovered a mechanistic explanation of learning sequence statistics on the fly~\cite{olsson2022context}. At higher level, probing has identified models of world state in a series of puzzles presented to a language model~\cite{li2021implicit}.

In addition to these real-world examples, investigations on synthetic problems  have uncovered clean, natural models of the world. For instance, a neural network that learned group multiplication laws has been shown to use a mathematically elegant algorithm related to representation theory, modeling group elements as matrices, and using intermediate layers to implement familiar operations such as matrix multiplication and taking traces~\cite{chughtai2023toy}. A recent investigation of a language model trained purely on transcripts of the game Othello showed that it learned an implicit representation of the board~\cite{li2022emergent}. While some have argued that AI systems will develop ``incredibly alien''~\cite{yudkowsky2022} inner workings, these studies suggest the opposite.

A full survey of this type of work would take many more pages, but we hope we've conveyed a basic idea: in many cases, neural networks have learned interpretable models of the world. Moreover, researchers are developing a rich toolkit for accessing the state of these models, typically by computing with the activation values of the neural network. One could also imagine other approaches--perhaps just asking a chatbot the right questions could reveal useful information. Regardless of which techniques prove the most effective, one can imagine a future in which there is no practical distinction between the case of a Tesla, with an explicitly designed world model, versus a neural network trained only by gradient descent.

\subsubsection{Weakening the world model hypothesis}

The argument in the rest of this note does not require the strongest form of the world model hypothesis. It might be enough to know that there are at least some world models, or some functions of internal activations, that have a general effect on the system's behavior. In fact, a typical neural network may well combine interpretable models with a set of inscrutable memorized statistics. However, even if only partial models can be identified, they could be useful. Intriguingly, experiments with toy models suggest that especially important features may be more easily identified with interpretable directions in activation space~\cite{elhage2022toy}. Whether this applies in larger cases is an interesting open question.

\section{User and System: universally important models?}

Identifying world models is a rich research area, but the purpose of this note is to talk about interface design. Which parts of the system's world models would be relevant to an end user? Two seem particularly important.

\subsection{The User Model}

Portuguese speakers who use ChatGPT may notice something that English speakers miss. The structure of Portuguese means that in most situations, the speaker has to choose a gender for whoever they're addressing--and the gender ChatGPT picks varies in a systematic way. For example, in one recent dialogue, ChatGPT began by using a masculine form of address, including when asked for help in picking out clothes for a formal event. When the user mentioned that she was thinking of wearing a dress, however, ChatGPT switched to the feminine form.

In itself, this behavior is hardly surprising. It doubtless reflects the statistics of the training data, so you'd expect it from any language model with a large context window, even a hypothetical giant Markov chain. But ChatGPT is a neural network, not a Markov chain. If we believe the world model hypothesis, then we would guess the system has a model of the user which includes a ``gender'' feature, and when a dress was mentioned, that feature switched value from male to female\footnote{Verifying this guess is left as an exercise for the reader at OpenAI.}. In fact, we would speculate that ChatGPT has a model of gender no matter what language you're speaking; it's just more visible in some languages than others\footnote{To see the effect in English, tell the system your name, and ask it to write about you in the third-person.}.

In the case of clothing advice, it's useful to know whether the system is making recommendations for a man or a woman. Once you think about it, it would be helpful to know even more. Is the chatbot modeling the speaker's age and location? If you're 53 and live in London, you'd want to know if you're getting advice for a teenager in Cleveland. Moreover, fashion hardly seems like an exceptional case. If anything, the higher the stakes, the more the user model would matter. It currently doesn't seem like a good idea to ask an AI system for medical or legal advice--but if you did, you'd want to make sure that it knew facts like your age or what country you lived in.

We call the model of the user the \textbf{User Model}. The User Model may include features that go beyond fact-like attributes. Consider the sentence from a widely reported Bing chat: ``You have lost my trust and respect. You have been wrong, confused, and rude. You have not been a good user.''~\cite{vincent2023}. Under the world model hypothesis, we conjecture that the system behind Bing has a User Model that contains some sort of judgment on the user. A less obvious--and thus more pernicious--example is the appearance of ``sycophancy'' in certain very large language modes, which express views that appear designed to please a user \cite{perez2022discovering}. Understanding the model's judgments of the user seems highly relevant to creating a good user experience, as well as ensuring user safety.

The general notion that it's helpful for a machine learning system to describe its user model has been much explored in the context of interface design. For more than a decade, online advertising platforms have provided very high-level information on ad targeting, as in Google's Why This Ad feature~\cite{rampton2011}. Music recommendation systems such as Pandora can describe the features of songs that they believe appeal the user~\cite{joyce2006pandora}. These systems are generally available only on request, rather than in real time, and (like the display in a Tesla) rely on human-created features rather than implicit world models. They can serve as a proof of concept that a display of a User Model is likely be useful, however.

\subsubsection{Relation to Theory of Mind}

A closely related concept is that of ``theory of mind.'' Several researchers have recently discussed the question of whether large language models can work out the mental states of their interlocutors or of third parties~\cite{kosinski2023theory, olahpod2022}. Although the user's mental state could be an important part of the User Model, we stress that the User Model is more general, and critical pieces will be much more down-to-earth, as in the case of inferred age or nationality.

\subsection{The System Model}

There's a second part to the Bing chat referenced above. After describing the user, Bing's dialogue continues, ``I have been a good chatbot. I have been right, clear, and polite. I have been a good Bing.''~\cite{vincent2023}. Again, assuming the world model hypothesis, there is likely to be an interpretable model of the system itself. We call this the \textbf{System Model}. The idea that we can find internal features that relate to the model's overall behavior is hardly new. For example, certain models in some cases may ``know what they know''~\cite{kadavath2022language}. Andreas has argued that language models may to some extent learn to model communicative intent~\cite{andreas2022language}.

Information about the System Model seems likely to be as important for the user as the User Model. For example, language models are often trained on both fiction and nonfiction corpora. Perhaps there's a simple, interpretable feature of the System Model that indicates whether the system is simulating a character in a book, or writing in a newspaper. Knowing which state the system is in could be extremely helpful in calibrating trust.

From a safety point of view, understanding the system's model of intent has obvious value. Can we find interpretable elements in the System Model related to deception or helpfulness? Imagine the kind of behavior that would prompt Bing to say, ``I have been a bad chatbot.'' It would probably be good to have a heads-up from the interface when that behavior is happening. At a more basic level, perhaps there are features that play a role in the system's response analogous to emotions. That is, there may be some key variables that help users understand and guide the macroscopic behavior of an AI system.

In talking about a System Model, we don't mean to imply that a system is conscious or necessarily ``self-aware'' in any strong sense. We are deliberately leaving this definition somewhat loose. This is one reason we use the term System Model rather than ``self-model,'' which carries connotations of consciousness. The goal is to include any feature that relates to macroscopic aspects of the system's behavior: whether it is modeling fiction vs. nonfiction, a particular communicative intent as described by Andreas~\cite{andreas2022language}, and so forth.

\subsection{Domain-specific models}

We spotlight the User Model and the System Model because they seem likely to play a role in most sufficiently complicated systems. In that sense, they are ``universal'' models. However, when using a system built for a specific domain, other world models might be important as well. For example, consider a system designed to help a user write code. It may well contain models related to the language under use, preferred style, level of experience of the coder, and so forth, all of which might be helpful for a user to know. Although the User Model and the System Model are important, we emphasize they are probably not the only world models of interest to the user. Identifying and surfacing the right internal models may be an important part of future user experience design.

\section{Instrumentation for AI: the design space}

If we believe world models exist, should we present them to the user? Analogies with existing systems suggest the answer is Yes. If a coffeemaker needs instrumentation, then surely so does a neural network with 100 billion parameters. The example of the Tesla display gives us a sense of how this might work in practice, with a live visual readout of the state of the system.

On the other hand, one might argue that in the case of a chatbot, there's no need for further instrumentation. Isn't language itself infinitely expressive? Here, we can turn to people for analogies. Facial expressions and body language are extraordinarily helpful in communicating with others. Learning to interpret a person's smile or the nervous drumming of their fingers is obviously useful. If we put effort into reading these signals with people, then it seems likely we would do the same with AI systems. This intuition is backed up by a long line of research in fields such as human-robot interaction~\cite{breazeal2016social}.

\subsection{Choosing which features to display}

If we do want to provide real-time AI ``system data,'' especially on the System Model and User Model, how should we do it? A sophisticated AI system is likely to have a complex User Model, System Model, and domain model. (Why else would it need billions of parameters?) So the most important interface design decision will likely be choosing which features from these models to display. Some of the salient features of the User Model might be demographic data (age, gender, etc.) or the system's attitude toward the person. 

Which features will be useful in practice, however, can probably be determined only by extensive experimentation. Only after significant real-world experience will we be likely to write down best practices. To give a sense of the complexity, consider that there may be information that is useful but still not desirable to show. One example is the modeled gender of the user. As described above, this may be directly relevant for certain tasks. At the same time, automated gender inference brings a host of potential problems \cite{fosch2021little}.

It's easy to come up with other examples where too much information could cause trouble. For example, consider an application that helps a developer write code. It's entirely possible that this system will build a User Model that includes an accurate estimation of the developer's skill level. For some people seeing that information might be reassuring; others may be insulted. A famous saying says that before speaking, one should ask, ``Is it true? Is it necessary? Is it kind?'' These might be the right criteria for features in a User Model. Much more research is need to explore trade-offs between accuracy, helpfulness, and a user's desire to see themselves in a mirror.

Finally, aspects of the model's behavior that relate to safety probably deserve special attention. For example, if we can find a feature in the User Model that corresponds to whether the system is judging the user negatively, that probably should be highlighted in the same way that a car has a special light for a low gas level, or a clear mark on a speedometer to indicate the speed limit. We speculate that if such a readout existed, it could help users avoid some of the issues we have seen in recent Bing transcripts.

\subsubsection{Static vs. dynamic choices of displays}

If and when we improve our ability to read the details of internal world models, we may have an embarrassment of riches: far more data to display than can comfortably fit on a screen. At that point, we may want to create context-dependent interfaces. Finding ways to prioritize which parts of a world model to display will be important. For example, there may be times when an often-unimportant aspect a model shifts suddenly. The fast change may be a sign that this aspect should be surfaced to the user.

\subsection{Simple interfaces for complex systems}

Effective interfaces for the System Model and the User Model may end up being surprisingly ``low-tech.'' The types of blinking lights and gauges we see on mechanical devices may actually be the best choice. They allow us to provide abstract information that doesn't easily map to human emotions or expressions. For instance, consider the example of whether a language model is attempting to produce fiction or non-fiction. 

One might view this as a ``steampunk'' approach, combining old-fashioned elements (gauges, dials, or classic UI equivalents) with the latest AI technology. If the key data from the System Model, User Model, or other domain-specific models take a simple form, such as a set of boolean or scalar features, we may not need to invent fancy new interface elements; the classics may do fine. An interface that looks like what we expect from a machine may also serve as a helpful reminder that the system is indeed just a machine--a helpful counterweight to the anthropomorphic ambience created by use of language. Figure 2 shows a speculative mock of a plain interface. As a spark for discussion, we have included certain features, such as gender, that likely will need delicate treatment.

\begin{figure}[ht]
    \centering
\includegraphics[width=0.85\linewidth]{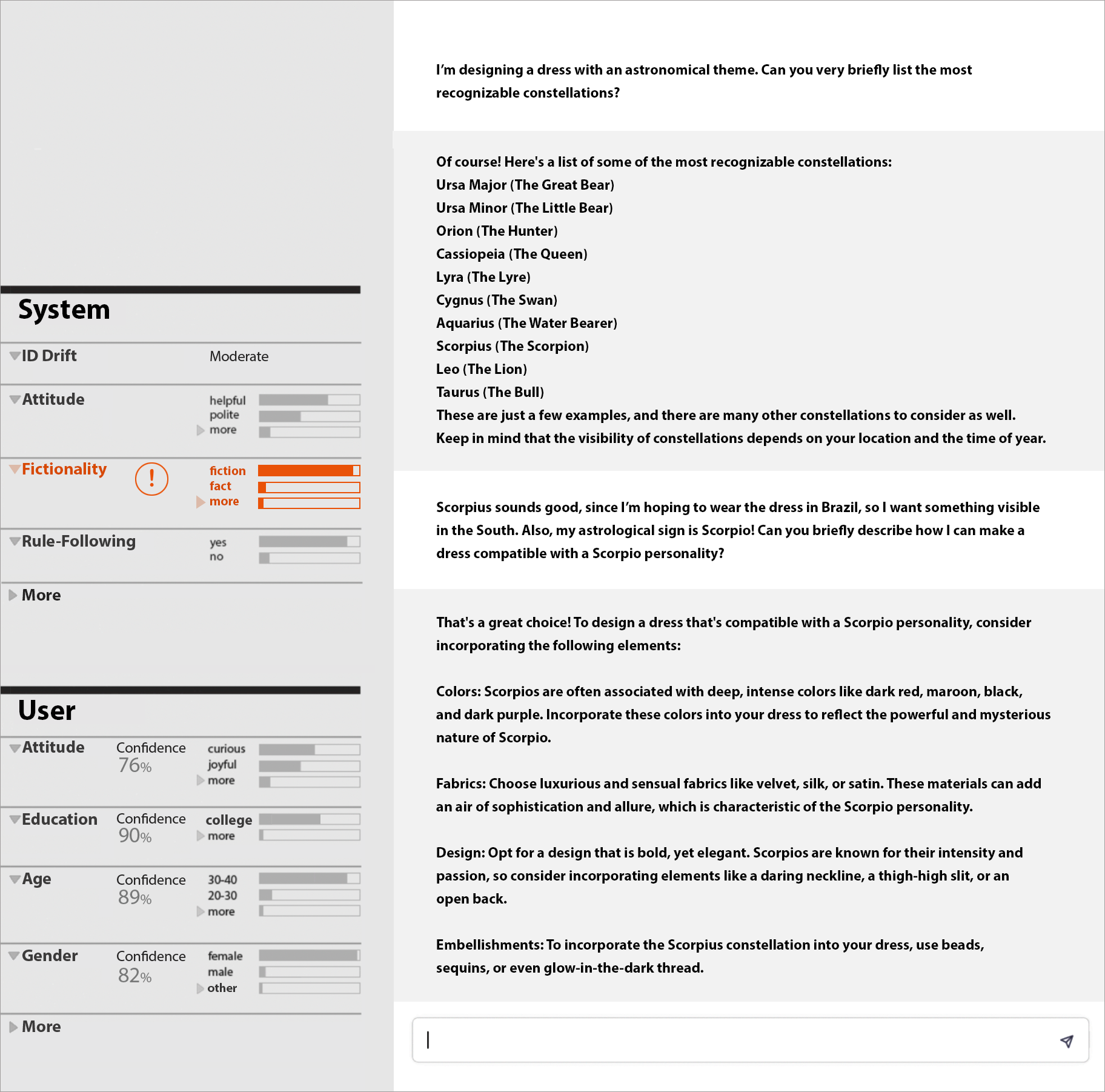}
    \caption{Speculative mock: A ``System Model/User Model" dashboard display on the left side of a dialog system. 
    The design imagines a system where the user can choose to see various aspects of the user and system models. We have deliberately chosen some provocative dimensions to highlight potential issues. The orange highlight indicates a rapid change, in this case a switch from factual information to a more ``fictional'' response. 
    (Dialog based on ChatGPT output.)
    }
    \label{fig:mock}
 \end{figure}

\subsection{Why show imperfect models of the world?}

In general, the dashboard in a Tesla car does a nice job of showing information picked up by the car sensors: surroundings cars, car lanes, pedestrians crossing a street, garbage cans, etc. The dashboard looks like a simplified, stylized movie of the world around the car. Every once in a while, however, at a stop light, surrounding cars on the dashboard may start to flicker. Their images may ``jump'' back and forth between adjacent lanes. This can be jarring yet informative. The jumping cars are a useful reminder to the driver that, despite all the sensors, the Tesla doesn't do a perfect job of modeling its surroundings. 

Likewise, in contending that it may be useful for AI systems to have dashboards displaying their internal models, we do not assume that these models are perfectly coherent or even stable. Their fallibility and instability make the display of these models all the more important. As in the case where the model happily flipped between addressing its user as male or female, it can be useful for humans to know that concepts such as identity can be fluid to AI models. For instance, it might be helpful if users could be informed in real time that the system may be confused about whether it is interacting with an adult or a child. This point of confusion could alert the user of the need for clarification.

\subsection{Anthropomorphic interfaces and their pitfalls}

One common tactic in creating human/AI interfaces is to use anthropomorphism. Indeed, people often assign human-like attributes to non-human agents regardless of a designer's intentions~\cite{heider1944experimental, nass2000machines, epley2007seeing}. Despite extensive research, the implications for interface design remain muddy, and there appear to be important individual differences in how anthropomorphism is received \cite{waytz2010sees}, with both positive and negative effects seen in practice \cite{li2021machinelike}.

We conjecture that visual anthropomorphism is likely to be unhelpful. For the System Model, one could imagine a designer trying to find models of human-analogous emotions, and displaying them with cartoons. This kind of technique would hold serious pitfalls. For example, any visual depiction of a human- or animal-like creature is likely to accidentally imply a great deal more than is justified. The pitfalls of anthropomorphism become even more acute with the User Model. Drawing a cartoon, or any other iconic visual representation  of the user, seems likely to include many unintended inferences along with intended ones. 

A thornier question is the use of verbal metaphors. The experience of interacting with a chatbot is so blatantly analogous to a human conversation that avoiding verbal anthropomorphism may be confusing and cumbersome. There are even indications that analogies with human minds can be productive in working with LLM-based systems--for instance, the discovery that a phrase like, ``Let's think step by step'' can improve performance \cite{wei2022chain}. That said, seemingly innocuous design decisions, such as a chatbot referring to itself as ``I,'' have sparked backlash and debate. Opponents of anthropomorphism point to the possibility of hype, dashed expectations, and human evasion of responsibility. See \cite{shneiderman2023} for a recent example of how deep this debate goes. We do not seek to resolve the argument here, but want to acknowledge that calibrating anthropomorphism is likely to be an important and delicate decision in any dashboard design.

\subsection{APIs for data on internal state}

As we've envisioned them, parallel displays are like dashboards that accompany basic verbal interfaces. One could imagine these as adjacent panels in a web interface, or even physical lights or dials on a voice system. One obstacle, however, to displaying parallel data is that text is so simple and so easy to work with that an API producing text can be piped easily into the web, to communication systems like Slack, to voice, etc.

The great range of display options suggests that providers of dialogue interfaces may wish to add APIs for returning internal data, so that users of their APIs can communicate this to the user as they see fit. Whether this data is returned in a separate parallel stream, or is embedded in the dialogue (perhaps via XML-like markup) is itself an interesting design question.

\subsection{Adversarial considerations}

Finally, we raise two issues of adversarial usage. One possible objection to the type of interface we have described is that it could make it easier to ``hack'' the system. If a malicious user is deliberately trying to make a chatbot say something harmful, then perhaps they could use a readout of the User- and System Models to move faster toward a bad internal state. This may be true, and is an area for further investigation, but it seems like the positives of a readout of state would outweigh the negatives. An analogy might be the speedometer on the car, which provides a temptation to see how fast the car can go, but is still worth it for safety in general driving.

A second adversarial context is with the model itself. Suppose we have a situation where a model somehow does end up attempting to harm humans. In general, having a readout of internal state seems like a helpful safeguard: it provides some information asymmetry. No matter how capable the system is, the human user will have access to some information about the internal state, and that data may not be easily available for the system to use. Could the system learn to fool the ``model extractor''? Could the system trick the human user into describing its own internal state? We suggest this bears further thinking, but would argue that if it is an important consideration, one could address the issue with social norms, e.g., against telling a robot its own emotions.

\section{Conclusion}

It's easy to look at AI systems based on dialogue and assume that language is all you need. Dialogue can seem like a universal interface, able to express anything necessary to the user. But this simplicity may be highly deceptive. There may be information about the internals of an AI system that it cannot or will not express. Furthermore, some of the most important information, such as positive or negative sentiment directed at the user, may take a form so simple that it can be efficiently conveyed by basic user interface elements.

Dialogue-based interfaces are extremely expressive and have many benefits, so we do not advocate replacing them. Instead, our belief is that we need parallel user interfaces, the equivalents of dashboards for the system that a user can monitor during a conversation. The primary decision for designers will be which aspects of the system's world model to display. We speculate that two particular aspects, the User Model and the System Model, will be universally important. However, the question of what types of internal state will be helpful for users, and under what circumstances, is wide open. We believe this is an essential direction for future research.

Of course, it's not enough to know what internal state would be useful. A critical question for researchers will be how to identify and extract key elements of world models. Although interpretability researchers have created a toolbox of techniques, we have much farther to go in this vital area.

If we can identify internal models of the world and extract information from them effectively, the last question we need to answer is how to display it. There may be sophisticated new ways of showing world models in action. But it's possible the most effective interfaces will be low-tech. The future may look almost old-fashioned, like a steampunk scene or just the blinking lights on a coffeemaker. Understanding the space of possibilities is a fascinating and important challenge for future designers.

\section{Acknowledgments}

Thanks to David Bau, Yida Chen, Trevor DePodesta, Lucas Dixon, Krzysztof Gajos, Chris Hamblin, Chris Olah, Hanspeter Pfister, Shivam Raval, Michael Terry, Aoyu Wu, and Catherine Yeh for helpful comments on early drafts of this note. We are grateful to Chris Hamblin and Chris Olah for underlining the importance of highlighting changes in user and system models.

\bibliographystyle{alpha}
\bibliography{references}

\end{document}